# Joint multi-contrast Variational Network reconstruction (jVN) with application to rapid 2D and 3D imaging

**Author list:**
Daniel Polak[1,2,3] Stephen Cauley[2,4,5], Berkin Bilgic[2,4,5], Enhao Gong[6], Peter Bachert[1,7], Elfar Adalsteinsson[8], Kawin Setsompop[2,4,5]

[1]Department of Physics and Astronomy, Heidelberg University, Heidelberg, Germany.

[2]Department of Radiology, A. A. Martinos Center for Biomedical Imaging, Massachusetts General Hospital, Charlestown, Massachusetts, USA.

[3]Siemens Healthcare GmbH, Erlangen, Germany.

[4]Harvard Medical School, Boston, Massachusetts, USA.

[5]Harvard-MIT Health Sciences and Technology, Massachusetts Institute of Technology, Cambridge, Massachusetts, USA.

[6]Subtle Medical Inc, Menlo Park, California, USA.

[7]Medical Physics in Radiology, German Cancer Research Center (DKFZ), Heidelberg, Germany.

[8]Department of Electrical Engineering and Computer Science, Massachusetts Institute of Technology, Cambridge, Massachusetts, USA.

**Word count**: 4154

**Acknowledgements**: This work was supported in part by NIH research grants: R01EB020613, R01EB019437, R01 MH116173, P41EB015896, U01EB025162, the shared instrumentation grants: S10RR023401, S10RR019307, S10RR019254, S10RR023043, and NVIDIA GPU grants.

# Abstract

**Purpose**: To improve the image quality of highly accelerated multi-channel MRI data by learning a joint variational network that reconstructs multiple clinical contrasts jointly.

**Methods**: Data from our multi-contrast acquisition was embedded into the variational network architecture where shared anatomical information is exchanged by mixing the input contrasts. Complementary k-space sampling across imaging contrasts and Bunch-Phase/Wave-Encoding were used for data acquisition to improve the reconstruction at high accelerations. At 3T, our joint variational network approach across T1w, T2w and T2-FLAIR-weighted brain scans was tested for retrospective under-sampling at *R*=6 (2D) and *R*=4x4 (3D) acceleration. Prospective acceleration was also performed for 3D data where the combined acquisition time for whole brain coverage at 1 mm isotropic resolution across three contrasts was less than three minutes.

**Results**: Across all test datasets, our joint multi-contrast network better preserved fine anatomical details with reduced image-blurring when compared to the corresponding single-contrast reconstructions. Improvement in image quality was also obtained through complementary k-space sampling and Bunch-Phase/Wave-Encoding where the synergistic combination yielded the overall best performance as evidenced by exemplarily slices and quantitative error metrics.

**Conclusion:** By leveraging shared anatomical structures across the jointly reconstructed scans, our joint multi-contrast approach learnt more efficient regularizers which helped to retain natural image appearance and avoid over-smoothing. When synergistically combined with advanced encoding techniques, the performance was further improved, enabling up to R=16-fold acceleration with good image quality. This should help pave the way to very rapid high-resolution brain exams.

# Introduction

Fast imaging techniques have been widely adopted into clinical practice to speed up MRI scans and thus help improve patient throughput, reduce the sensitivity to involuntary patient motion [1], improve patient compliance and potentially obviate the need for sedation in pediatric patients [2]. However, conventional parallel imaging (PI) algorithms (e.g. *SENSE* [3], *GRAPPA* [4], etc.) are constrained to moderate acceleration rates, *R*, (e.g. typically *R*=3 for 2D and *R*=2x2 for 3D) to avoid structural artifacts and large noise amplification. To enable higher accelerations with improved image quality, advanced encoding and reconstruction techniques have been proposed.

Among these techniques, *2D-CAIPIRINHA* [5] is applicable to volumetric 3D acquisitions and employs a staggered $k_y$-$k_z$ under-sampling pattern to create controlled aliasing in the phase (y) and partition (z) encoding plane which increases the distance between the aliasing voxels and enables better utilization of coil sensitivity information in the reconstruction. *Wave-CAIPI* [6] adopts this scheme and combines it with *Bunch Phase Encoding (BPE)* [6] by playing additional sinusoidal gradients on both the $G_y$ and $G_z$ gradients with a quarter-cycle phase shift during the readout. This enables controlled aliasing along all three spatial dimensions, including the readout axis (x), which significantly reduces artifacts and g-factor noise amplification when compared to *2D-CAIPIRINHA*. At 3T, the *Wave-CAIPI* technology was demonstrated to provide up to 9-fold acceleration for 3D sequences [7]–[9] with comparable diagnostic quality as *GRAPPA* at the lower acceleration rate of *R*=4. Moreover, Wave-CAIPI was employed in Simultaneous Multi-Slice (*SMS)* sequences [10], where an effective acceleration of *R*=12-fold was achieved as multiband (MB) acceleration does not cause $\sqrt{R}$-SNR penalty. However, in general the efficiency of this technique is significantly reduced when applied to 2D sequences (without *SMS*) where controlled aliasing is limited to the x-y domain (cf. *BPE*). Besides artifacts also SNR can be a challenge at very high acceleration (both for 2D and 3D) due to the inherent $\sqrt{R}$-noise penalty and may necessitate going to higher magnetic field strength (as in *Wave-GRAPPA* [11]) or using frameworks like *Compressed Sensing (CS)* [12] and *LORAKS* [13], which have also been synergistically combined with *Wave-CAIPI* [14], [15].

However, for techniques like *CS* to work robustly, several pre-requisites need to be fulfilled. Aliasing artifacts must be incoherent which is commonly achieved by non-Cartesian or random under-sampling, but since most clinical sequences employ Cartesian sub-sampling, incoherence is in practice limited to dynamic- and 3D sequences but remains a challenge for 2D acquisitions. Moreover, *CS* requires the existence of a representation in which the reconstructed images become sparse. Commonly used transformations are wavelet [12], TV [16] and TGV [17] which in combination with the $\ell_1$ norm achieve at least approximate sparsity. However, the use of the $\ell_1$ norm entails iterative optimization algorithms which are often computationally demanding and yield longer reconstruction times. Also, the choice of the regularization parameter(s) is crucial to prevent over-smoothing.

Recent developments in deep learning have the potential to lift some of these barriers. On highly accelerated data, neural networks have outperformed existing techniques both in terms of image quality, artifact reduction as well as reconstruction time. The algorithms proposed in [18] and [19] operate on coil-combined images and were trained to un-aliase zero-padded reconstructions or enhance the image quality of conventional methods such as *SENSE*, *GRAPPA* or *CS*, etc. Moreover, further improvement was demonstrated by reconstructing multiple clinical contrasts jointly. This idea was previously investigated for *PI+CS* reconstructions where additional sparsity constraints along the contrast dimension were used [20]–[22] and this concept has now also been applied to deep learning [23]. By exploiting the redundancy across the jointly reconstructed contrasts, these techniques enable better image quality than single-contrast methods. However, the pixel-wise loss used in these approaches requires the multi-contrast data to be spatially registered, which may pose a challenge for clinical routine. A recent work [24] discovered the relevance of this issue and proposed a conditional GAN with cyclic consistency loss [25] to jointly reconstruct unregistered multi-contrast data.

Several groups have demonstrated the benefits of incorporating the multi-channel MRI data into the deep learning reconstruction. *RAKI* [26] is a k-space based technique where a convolutional neural network (CNN) is trained to synthesize non-acquired lines in k-space. When compared to *GRAPPA*, which is a linear interpolation method, this non-linear extension yields improved noise resilience at high acceleration. Also, *RAKI* may be favorable as the training is performed solely on the subject specific ACS data and hence large amounts of training data are not required. *AUTOMAP* [27] takes this one step further by learning the entire transformation from under-sampled multi-channel k-space data to the final image without ever explicitly using the Fourier transformation. This may present a flexible alternative for the reconstruction of non-Cartesian k-space trajectories where the exact inverse transform may not exist.

Inspired by traditional iterative techniques for inverse problems, several approaches [28]–[30] have posed the MRI image reconstruction as an unrolled gradient descent optimization where the physics model is embedded in the reconstruction and regularizers/priors are learnt from training data. This formulation can be understood as a generalization of *CS* where neural networks are utilized instead of hand-crafted domain transformations (such as wavelet or TV). With this framework many existing physics- and *CS*-based techniques have been outperformed while enabling much shorter reconstruction times [28]. In a recent work [31] such a network was also utilized to reconstruct a highly accelerated *Wave* acquisition where imperfections of the sinusoidal *Wave* gradient trajectory were automatically estimated by the network without additional time-consuming optimizations (e.g. AutoPSF [32]).

In this contribution, we augment the unrolled gradient descent optimization in the variational network (*VN*) architecture [28] to jointly reconstruct multiple clinical contrasts (T1w, T2w, T2-FLAIR) from accelerated MRI acquisitions. By taking advantage of the shared anatomical information across the jointly reconstructed scans, our joint variational network (*jVN*) approach learns more efficient regularizers which improved the image quality when compared to single-contrast *VN* reconstructions. Moreover, we investigated how complementary k-space sampling across imaging contrasts and advanced acquisition techniques such as *BPE* and *Wave*-encoding can be utilized to further boost the reconstruction performance. We validated these techniques both on 2D and 3D data and ultimately demonstrate the feasibility of obtaining T1w, T2w and T2-FLAIR contrasts at 1 mm isotropic resolution with $R$=16-fold acceleration in less than three minutes of scan time.

## Methods

### Network architecture of jVN

This work is based on the variational network (*VN*) architecture [28] which aims to solve the PI problem as an unrolled gradient descent (GD) optimization (Fig. 1), where each step contains non-linear filtering and data-fidelity operations. Our joint variational network (jVN) augments this technique by reconstructing multiple clinical contrasts simultaneously. This is achieved by stacking e.g. T1w, T2w and T2-FLAIR-weighted images along the channel dimension of the network. Starting from an initial reconstruction $\vec{u}^0$, each gradient descent step mixes the $N_c$ input contrasts by convolving them with the filter kernels $k^t$ resulting in $N_k$ feature channels. Next, learned activation functions $\phi^t$ and the transposed filter $\bar{k}^t$ are applied to reduce the number of feature channels to the number of input contrasts. Moreover, a data-fidelity term $A_c^H(A_c u_c^t - f_c)$ weighted by a trained regularization parameter $\lambda_c^t$ is computed for each contrast individually (no mixing between scans) and subtracted from $\vec{u}^t$ at the end of each step.

As explored by previous contributions [33], the image quality of multi-contrast reconstructions can be improved using complementary k-space under-sampling, e.g. by varying the acceleration factor across the input contrasts and/or including a contrast-dependent shift in the k-space sampling. This results in different aliasing and image artifacts in the initial *SENSE*-based reconstructions of the different image contrasts, which can be leveraged in multi-contrast reconstructions. In this work, we chose to keep the acceleration factor fixed but shift the uniform under-sampling pattern (see Fig. 2) for each contrast. Particularly in 3D, this approach simplifies the data handling as the coupled voxel locations are identical across all contrasts; note, that we use *jVNc* to refer to joint variational network reconstructions with complementary k-space sampling.

Moreover, we utilized *BPE* to improve the quality of our 2D scans and *Wave*-encoding for our 3D acquisitions. Since these techniques couple the readout dimension (x) into the PI problem, the encoding matrix is no longer separable along this dimension- and reconstructing a full dataset at once may be intractable on state-of-the-art GPUs especially for high-resolution 3D scans. To mitigate this issue, we constrained our acquisitions to uniformly under-sampled k-space masks with fixed acceleration. This allows the PI reconstruction to be split into smaller sub-problems of collapsing voxels in image space which has the dimension $R_y$ x $R_z$ x $N_x$. To account for these adaptions, we modified the network's forward model operator $A_c = \sum_{z_j} \sum_{y_i} \mathcal{F}_x^{-1} \text{Psf} \mathcal{F}_x C p_c$ [7] where the index $c$ denotes the contrast dependency and $\mathcal{F}_x$ the Fourier transformation along x. The $A_c$ operator first applies a linear phase ramp $p_c$ to reflect shifts in the uniform k-space sampling mask and then multiplies with the coil sensitivity $C$ and *Wave* point-spread-function $\text{Psf}$ in hybrid space $[k_x, y, z]$. Ultimately, $A_c$ sums over the collapsing voxels $y_i, z_j$ ($i \in [1 \dots R_y]$, $j \in [1 \dots R_z]$) corresponding to the acceleration factors $R_y$ and $R_z$.

Furthermore, the following modifications were implemented: As no internal autocalibration scan (ACS) data was used for any of the reconstructions (i.e. no fully-sampled center of k-space), we generated input images $u^0$ from an initial *SENSE*-based instead of a zero-padded reconstruction (in contrast to [28]). We empirically observed that this improved the image quality for all evaluated reconstructions. Moreover, we trained individual networks for every output contrast, which was found to provide overall better image quality than a single network. We implemented this in the training stage by extracting one image contrast from the vector $\vec{u}^T$ (containing all jointly reconstructed contrasts) before minimizing the $\ell_2$-loss with respect to the corresponding ground truth data. In this way, the loss function only measured the fidelity in a single contrast instead of all the input images.

## Data acquisition and pre-processing

With IRB approval and informed consent, fully sampled training data were acquired on eight healthy subjects using two 3T scanners (MAGNETOM Prisma and Skyra, Siemens Healthcare, Erlangen Germany) and a product *SPACE* sequence (variable flip angle 3D Turbo Spin Echo [34]) with T1w, T2w and T2-FLAIR-weighted contrasts (FOV: 256x256x192 mm³, resolution: 1x1x1 mm³, orientation: sagittal, BW: 592Hz/px, product Siemens 32-channel head coil). As the training of our networks was solely performed on retrospectively under-sampled datasets with synthesized *BPE/Wave*, we also acquired accelerated data for prospective testing (on a separate subject). For these acquisitions at *R=4x4*, a prototype *Wave SPACE* sequence [9] with complementary k-space sampling (Fig. 2) and four sinusoidal *Wave* cycles per readout with 16 mT/m gradient amplitude was used. The combined acquisition time for T1w, T2w and T2-FLAIR was TA=2:53 min including a two second external *GRE* reference scan for the computation of coil sensitivity maps. The same contrasts were also acquired at *R=2x2* acceleration (without *Wave*-encoding) and were reconstructed using *SENSE*.

For the training and testing on retrospectively under-sampled datasets, all fully sampled 3D scans were first co-registered channel-by-channel using FSL FLIRT [35] to mitigate any inter-scan motion between the acquisitions of T1w, T2w and T2-FLAIR-weighted scans. For our 2D experiments, the registered images were then reformatted into axial datasets with 1 mm in plane resolution and 4 mm slice thickness (whole-brain coverage). Next, the central 20x20 lines of k-space were extracted from T2-FLAIR and coil sensitivity maps were computed using ESPIRiT [36]; only for the prospectively accelerated acquisitions an external GRE reference scan was used. *BPE* (for 2D) and *Wave*-encoding (for 3D) were synthesized by convolving the fully sampled datasets with a point-spread-function (Psf) corresponding to the *Wave* acquisition parameters described above. The training data was retrospectively under-sampled at *R*=6 for 2D and *R*=4x4 for 3D using the complementary k-space sampling scheme described in Fig. 2 and reconstructed using generalized *SENSE*. For the prospectively accelerated data acquired with *Wave*-encoding, imperfections of the sinusoidal *Wave* gradient trajectory were first estimated using AutoPSF [32] (entirely data-driven, no additional calibration scans). The resulting Psf was then utilized in the generalized *SENSE* and variational network reconstructions.

### Training and testing

To assess the benefit of reconstructing multiple contrasts jointly and/or utilizing complementary k-space sampling and *BPE/Wave*, separate networks were trained while the following parameters were held constant: T=10 iterations, $N_k$=24 feature channels, kernel size 11x11, learned activations from 31 radial basis functions [28]. For our 2D scans, 1008 axial slices from seven subjects were used for training (batch size: 5; epochs: 250), testing was performed on 36 slices from the remaining subject which was not used in the training.

We also characterized potential artifacts in the presence of inter-scan motion where the jointly reconstructed scans are not spatially registered and evaluated the performance of a preliminary motion correction technique. Inter-scan motion was simulated by applying in-plane translation Δs and/or in-plane rotation $\theta$ to the fully sampled T1w (Δs=(2,2)$^T$ mm, $\theta$=3°) and T2w (Δs=(-2,-2)$^T$ mm, $\theta$=-3°) test dataset before repeating the pre-processing as described above. Combined with the unchanged T2-FLAIR scan, this resulted in a multi-contrast dataset, where the jointly reconstructed scans were not spatially registered. We reconstructed this test dataset using our *jVNc+BPE* network that was solely trained on registered images and assessed potential artifacts. Moreover, we evaluated a preliminary correction technique where the image estimates $\vec{u}^t$ were registered in every iteration of the network. For this, additional translation and rotation operators were placed before the convolutional filter $k^t$ and corresponding inverse transformations after $\bar{k}^t$ to retain agreement with the acquired scanner data (note, that bilinear interpolation was used in all translation and rotation operations). We tested this setup using both the exact and estimated motion parameters that were obtained by registering the initial *SENSE* reconstructions $\vec{u}^0$ (MATLAB *imregister*).

For our 3D datasets at *R*=4x4, separate networks were trained with and without complementary k-space sampling and *Wave*-encoding using the same architecture as described for 2D. However, as each training sample now consisted of four sagittal slices ($R_z$=4), the batch size was reduced to two to limit the required GPU memory. Overall, 336 training samples (1344 slices) from seven subjects were used for training, while testing was performed on 48 samples (192 slices) from the remaining subject. Moreover, *VN* and *jVNc+Wave* were tested on prospective scans at *R*=4x4 where Wave-encoding was used in the acquisition and was not synthesized.

# Results

Figure 3 demonstrates the results for T2-FLAIR at $R$=6-fold acceleration. As shown, the encoding capability of *SENSE* was insufficient at such high acceleration, causing large noise amplification and residual aliasing artifacts. However, the single-contrast *VN* network mitigated most of these issues, but the artifact and noise reduction came at the cost of over-smoothing and loss of spatial resolution, as indicated by the zoom-in. When reconstructing T1w, T2w and T2-FLAIR contrasts jointly (*jVNc*) or utilizing BPE (*VN+BPE*), the NRMSE was decreased. Moreover, fine anatomical details were also better preserved as demonstrated by the improved conspicuity of a blood vessel in *jVNc* (thin arrow) or a better-defined region of CSF in the posterior of the brain for *VN+BPE* (bold arrow). However, the overall best performance was achieved when *jVNc* was synergistically combined with *BPE*. This is best seen in the zoom-in, where for example the gray-white matter boundary (bold arrow) or a small line of CSF (thin arrow) became visible which were over-smoothed in all other reconstructions. These improvements are also reflected in better NRMSE, SSIM and PSNR which are provided in Tab. 1.

Figure 4 displays the results for T1w, T2w and T2-FLAIR at $R$=6-fold acceleration. Across all imaging contrasts, *jVNc+BPE* better retained the spatial resolution when compared to *VN* which is best seen in the anterior part of the brain (bold arrows) where the gray-white matter boundary is over-smoothed. Moreover, the comparison demonstrates that the sequence-specific contrast was preserved without signal leaking from one scan to another. This is best seen in the center of the brain where two thin arrows mark a circular region of CSF and a blood vessel. Both have similar geometric shape and low signal intensity in T1 and T2-FLAIR, while in the T2w scan one of them (CSF) is hyper-intense. Despite the non-linear mixing of all input contrasts in the convolutional filters of the joint variational networks, no change in signal intensity was observed in any of the reconstructions, while the conspicuity of these anatomical features was much improved when compared to *VN* and *SENSE*.

In Figure 5, the effect of inter-scan motion in joint multi-contrast reconstructions was analyzed. Our *jVNc+BPE* network resulted in poor image quality with residual aliasing artifacts (see red arrows) when the input images were not aligned and NRMSE was even worse than the corresponding single-contrast *VN+BPE* reconstruction. Our preliminary motion correction technique efficiently mitigated such artifacts (green arrows) and provided similar image quality as observed without inter-scan motion. The underlying motion parameters were estimated using the initial *SENSE* reconstructions and were in good agreement with the exact parameters. While

NRMSE was almost entirely unaffected by translations, a slight increase was observed in the presence of rotations both when the exact and estimated motion values were used.

The results of the 3D reconstructions at $R$=4x4 acceleration are displayed in Fig. 6. Again, the *SENSE* reconstruction suffered from severe noise amplification from the $\sqrt{R}$ and g-factor noise penalty. In contrast, the *VN* network mitigated large noise enhancement, but the coronal reformats exhibit striping artifacts due to the ill-conditioning of the reconstruction which was performed sequentially across the aliasing coronal slice groups (convolutional filters in *VN* were applied to sagittal cuts). Moreover, the zoom-in reveals residual aliasing and loss of spatial resolution (bold arrow) in regions of high g-factor where the encoding capability of the 32-channel head coil is limited. In contrast, the *jVNc* network with complementary under-sampling helped to reduce some of these artifacts (thin arrow) and improved NRMSE, but the striping artifacts were only mitigated when the PI problem was better conditioned using *Wave*. The lowest NRMSE was obtained by *jVNc+Wave*, as demonstrated in the zoom-in, where the gray-white matter boundary was best preserved.

Finally, we tested our variational networks on prospectively accelerated data ($R$=4x4) acquired with and without *Wave*-encoding. The results are displayed in Fig. 7 and Fig. 8, where a conventional acquisition at $R$=2x2 acceleration (no *Wave*-encoding) served as the reference. Both variational network reconstructions were able to preserve the sequence-specific contrast, however *jVNc+Wave* more efficiently removed aliasing artifacts (thin arrow) and better preserved the spatial resolution (bold arrow). Nevertheless, at such high acceleration also *jVNc+Wave* suffered from slight image blurring, e.g. in the cerebellum of T2-FLAIR. Moreover, the results demonstrate that our networks generalized to prospective acquisitions although the training data was under-sampled retrospectively and *Wave*-encoding was synthesized.

## Discussion

In this contribution, we developed a framework to reconstruct data from multiple clinical imaging contrasts jointly using the variational network architecture. By utilizing shared anatomical information across the imaging contrasts, *jVN* learned more efficient regularizers, which enables the reconstruction of highly under-sampled datasets with significantly reduced artifacts and image blurring. Moreover, we incorporated advanced encoding techniques in our acquisitions and demonstrated the benefit of complementary k-space under-sampling and *BPE/Wave*-encoding. This allowed T1w, T2w and T2-FLAIR-weighted scans to be acquired and jointly reconstructed at *R*=6-fold acceleration for 2D and up to *R*=16-fold acceleration for 3D (combined TA<3 min), while retaining good image quality.

We quantitatively assessed the benefits from reconstructing multiple contrasts jointly and/or utilizing advanced encoding schemes such as *BPE* or *Wave* and showed that the synergistic combination yielded the overall best results. While the former technique allows the network to learn more efficient regularizers by leveraging shared anatomical structures across the jointly reconstructed contrasts, the latter improves the overall conditioning of the PI reconstruction by exploiting variations of the coil sensitivity also along the readout. It was observed that the combined approach enabled higher improvement for 2D compared to 3D, where at *R*=6-fold acceleration the standard *SENSE* reconstructions resulted in large residual aliasing due to the insufficient encoding capability. In contrast, the *R*=16-fold accelerated *SENSE*-reconstructions for 3D were less effected by artifacts but dominated by the $\sqrt{R}$-SNR and g-factor noise penalties ($\sqrt{R}$=4). This suggests that learning more efficient regularizers in *jVNc* mainly helps to resolve structural aliasing (as in our 2D scans) but is less beneficial in the presence of low SNR and few artifacts (such as in our 3D scans).

We also assessed the performance of *jVN* in the presence of inter-scan motion, where the different clinical contrasts were not spatially aligned. Such motion may occur e.g. between pre- and post-contrast acquisitions, where there is typically of delay of several minutes. Our preliminary simulation on 2D data revealed that spatial miss-alignment in joint multi-contrast reconstructions may result in poor image quality and potentially worse performance than the corresponding single-contrast *VN* reconstruction. This is intuitively clear, as the training was solely performed on registered data and the network learned to leverage this property as an additional prior in the reconstruction. However, such artifacts were almost entirely removed using our proposed motion mitigation technique which embeds translation and rotation operators in the network and utilizes the initial *SENSE* reconstructions to estimate the motion parameters. While good image quality was obtained on all test datasets, we observed a slight increase in NRMSE for rotations even when the exact motion parameters were used, which we assume is a consequence of the bilinear interpolation method used in this work. However, on the cost of some minor increase in computation time, further improvement is expected by employing more advanced interpolation techniques such as spline or sinc. Moreover, future work is required to analyze the sensitivity to errors in the motion estimation and how this could degrade the image quality of joint multi-contrast reconstructions. Although good agreement between the estimated and exact motion parameters was observed for the simulated in-plane translation and rotation, these results still need to be validated in in-vivo acquisitions, where further complications might arise from through-plane and intra-scan motion. In the latter case, prospective [37] or retrospective (data-driven) correction techniques [38], [39] could help to reduce associated artifacts.

Also, special attention was payed to potential artifacts caused by the mixing of clinical contrasts in the convolutional filters. While this was found to be beneficial for the reconstruction of highly accelerated datasets, it bears the risk of signal leaking from one scan to another, which could impede the clinical diagnosis. All test slices were carefully reviewed however such artifacts were never observed and we have the following explanation for this. In *jVN* multiple data-fidelity computations are embedded throughout the feed-forward path of the network which helps to hold the contrast-mixing in check. If signal leaked from one scan to another, the data inconsistency between the current estimate of the image and the acquired scanner data would increase. Also, due to the coupling in the PI problem, artifacts would not only remain at the location of origin but spread to all coupled voxel locations, a global penalty which the filters in *jVN* efficiently learnt to avoid. Nevertheless, due to the small number of test subjects available in this study, further investigation with larger patient cohorts is necessary to confirm these initial findings especially in the presence of pathology.

In this work, we trained *jVN* across three contrasts (T1w, T2w, and T2-FLAIR-weighted) which are commonly used in clinical brain exams. However, we anticipate that further acceleration feasibility can be achieved by increasing the number of clinical contrasts, e.g. by including T2*w (SWI) [7] and/or post-contrast T1w scans [8]. This would not only provide improved regularization from increased anatomical information but also enable more efficient complementary k-space sampling and could pave the way for a very rapid multi-contrast brain exam (cf. [9], [40]). However, in such undertaking, it will be important to assess and potentially refine the *jVN* architecture to enable robust reconstructions across imaging contrasts with large background phase differences. In this work, all scans were acquired using a 3D TSE sequence, which resulted in the same image phase across all contrasts. However, phase variations may also arise from the coil sensitivity maps themselves which in the development phase of this work were calculated from the fully sampled k-space data of each imaging contrast individually and then included in the forward model of the reconstruction. It was observed that such phase differences can cause degradation in the reconstruction performance when compared to reconstructions that use the same set of coil sensitivities across all the contrasts. We anticipate that such behavior is specific to the *VN* architecture, where real and imaginary feature maps are summed after the convolutional filtering ($k^t$) and are not being kept as separate channels (cf. [31]). Moreover, we expect further improvement in the overall image quality by switching to deeper network architectures which should benefit both the single- and multi-contrast reconstructions.

In conclusion, we demonstrated the benefit of reconstructing multiple clinical contrasts jointly and investigated how complementary under-sampling and *BPE/Wave*-encoding can be facilitated to improve the image quality. We carefully evaluated the performance of our networks both on 2D and 3D acquisitions, analyzed potential artifacts from inter-scan motion and finally demonstrated the feasibility of obtaining T1w, T2w and T2-FLAIR-weighted contrasts at high isotropic resolution in less than three minutes of scan time.

# Tables

|  | *SENSE* | *VN* | *jVN* | *jVNc* | *VN+BPE* | *jVN+BPE* | *jVNc+BPE* |
|---|---|---|---|---|---|---|---|
| NRMSE [%] | 17.95 | 10.29 | 8.69 | 6.58 | 6.63 | 5.88 | **5.00** |
| SSIM | 83.36 | 94.79 | 95.95 | 97.45 | 97.54 | 98.12 | **98.53** |
| PSNR | 31.56 | 36.39 | 38.00 | 40.28 | 40.26 | 41.42 | **42.66** |

Table 1: Quantitative metrics (NRMSE, SSIM and PSNR) are provided for the T2-FLAIR reconstructions at *R*=6-fold acceleration. Improvement over *VN* was achieved by either reconstructing all contrasts jointly (*jVN*), employing complementary under-sampling (*jVNc*) or utilizing *BPE*. The overall best results were obtained from the synergetic combination (*jVNc+BPE*) and are highlighted in bold.

# Figures

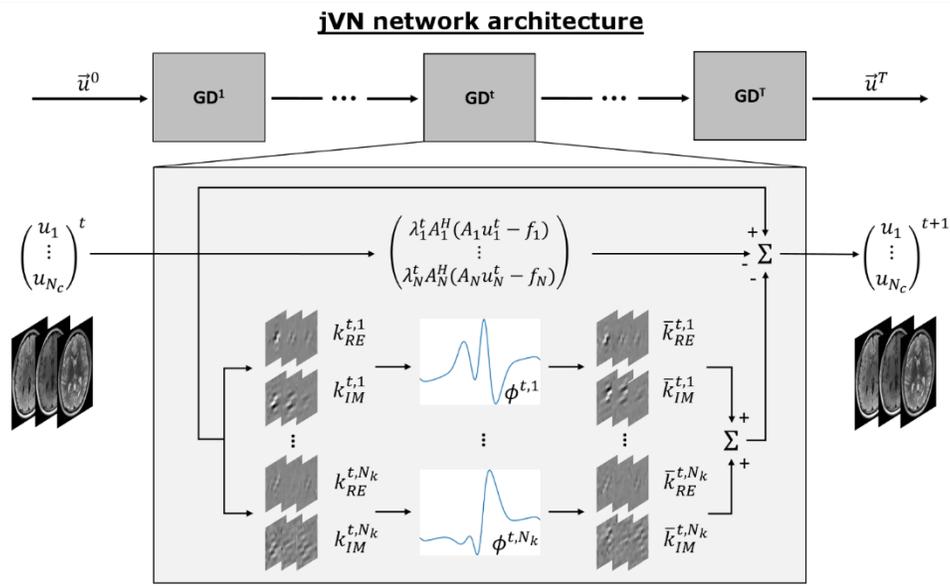

Figure 1: *jVN* is based on the variational network architecture [28] and poses the image reconstruction as an unrolled gradient descent (GD) optimization. Each gradient descent step GD$^t$ contains a convolutional filter $k^t$ which mixes the different input contrasts $\vec{u}^t$ and creates N$_k$ feature channels. Non-linear activation $\phi^t$ and the transposed filter $\bar{k}^t$ reduce the N$_k$ feature channels to the number of input contrasts N$_c$. Data-fidelity is computed individually for each contrast, where each forward model matrix A$_c$ contains a contrast-specific under-sampling mask that can vary between contrasts to enable complementary k-space sampling (compare Fig. 2). For *BPE/Wave* acquisitions, A$_c$ additionally contains the *Wave* point-spread-function (Psf) to account for the voxel spreading along the readout direction.

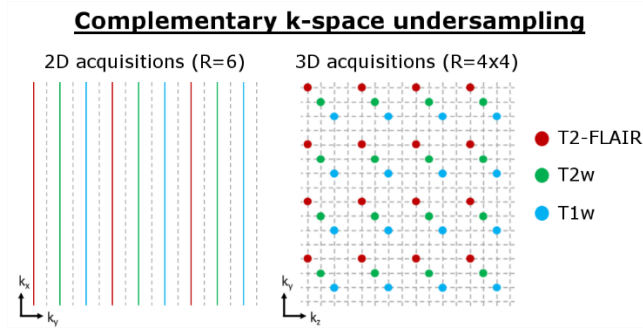

Figure 2: Our multi-contrast reconstructions (*jVNc*) employ complementary k-space under-sampling by imposing a contrast-dependent shift on the uniform sub-sampling mask.

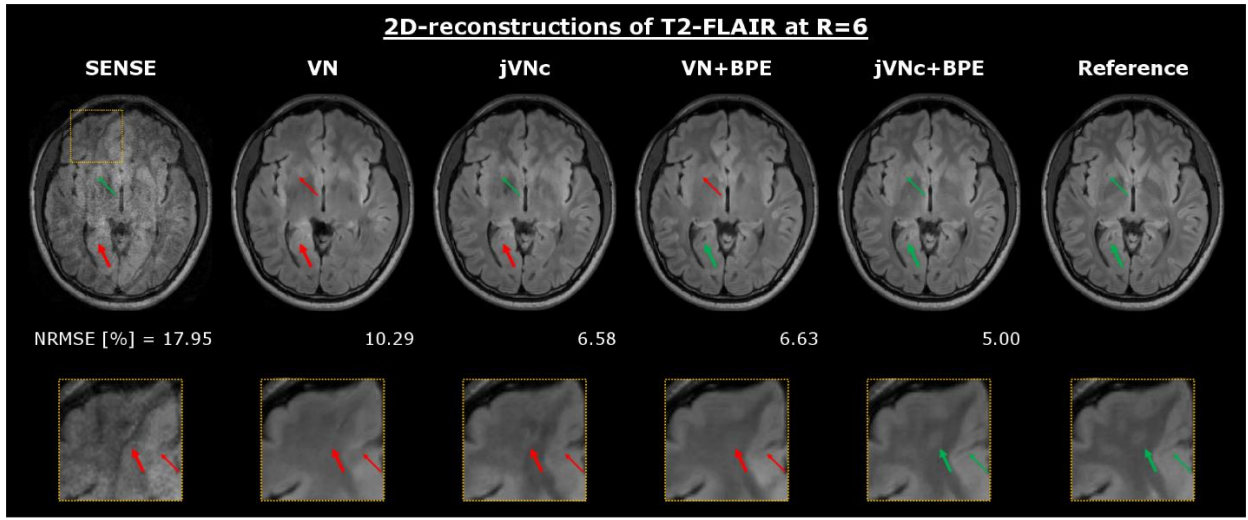

Figure 3: At *R*=6-fold acceleration, the *SENSE* reconstruction of T2-FLAIR resulted in large noise amplification and aliasing artifacts, which were mostly mitigated using the single-contrast *VN* network. However, by reconstructing T1w, T2w and T2-FLAIR contrasts jointly (*jVNc*) or utilizing BPE (*VN+BPE*), fine anatomical details were better preserved and the over-smoothing reduced when compared to *VN*. The overall best performance was achieved by *jVNc+BPE* which is also reflected in the lowest NRMSE.

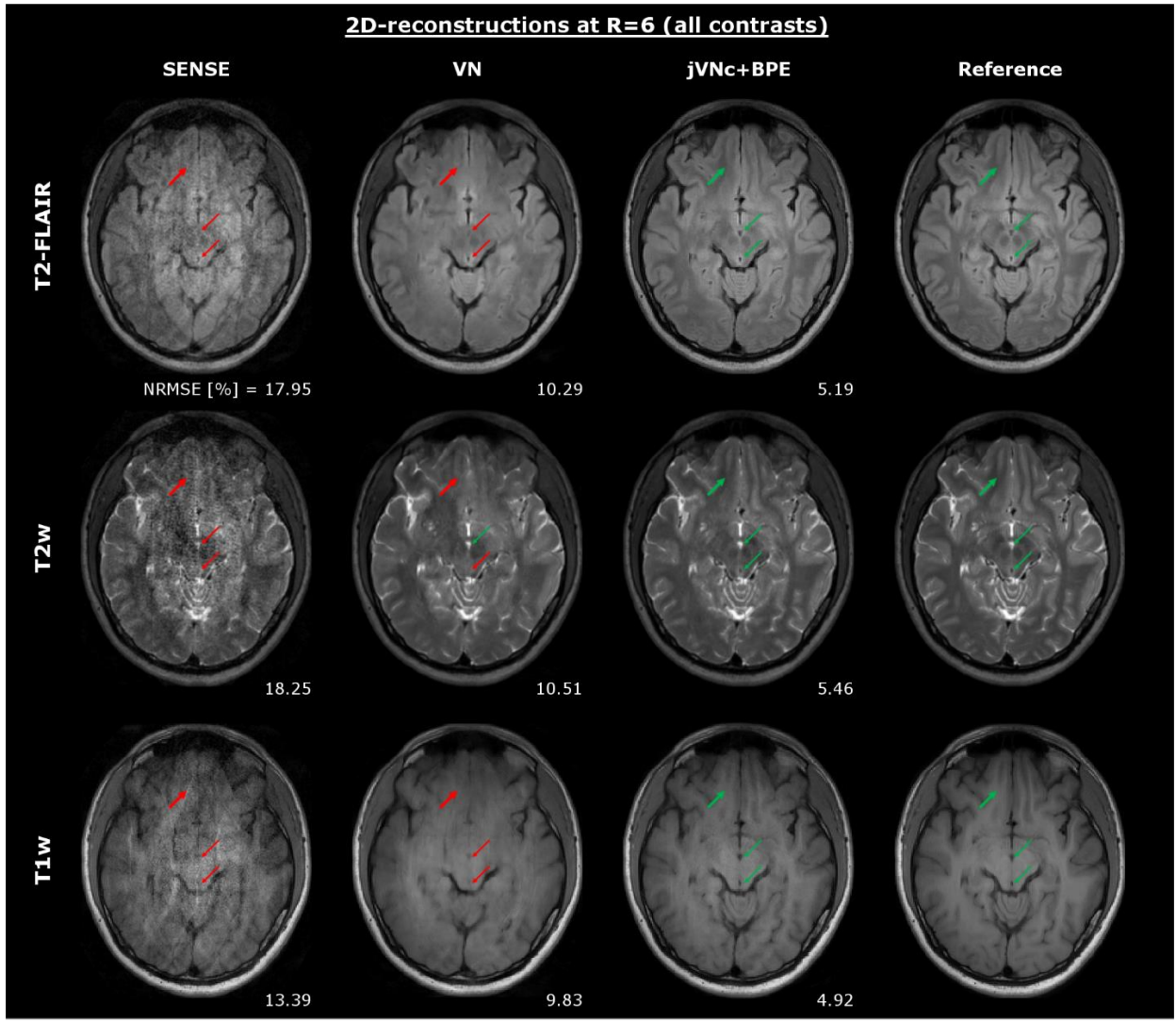

Figure 4: Throughout all contrasts, *jVNc+BPE* better preserved the spatial resolution (bold arrows) and achieved lower NRMSE compared to *VN*. Moreover, the comparison demonstrates that *jVNc+BPE* retained the scan-specific contrast (thin arrows). Signal leakage from one contrast to another was not observed, as exemplarily demonstrated for a blood vessel (dark in all contrasts) and a region of CSF (hyper-intense only in T2w, but dark in T1w and T2-FLAIR).

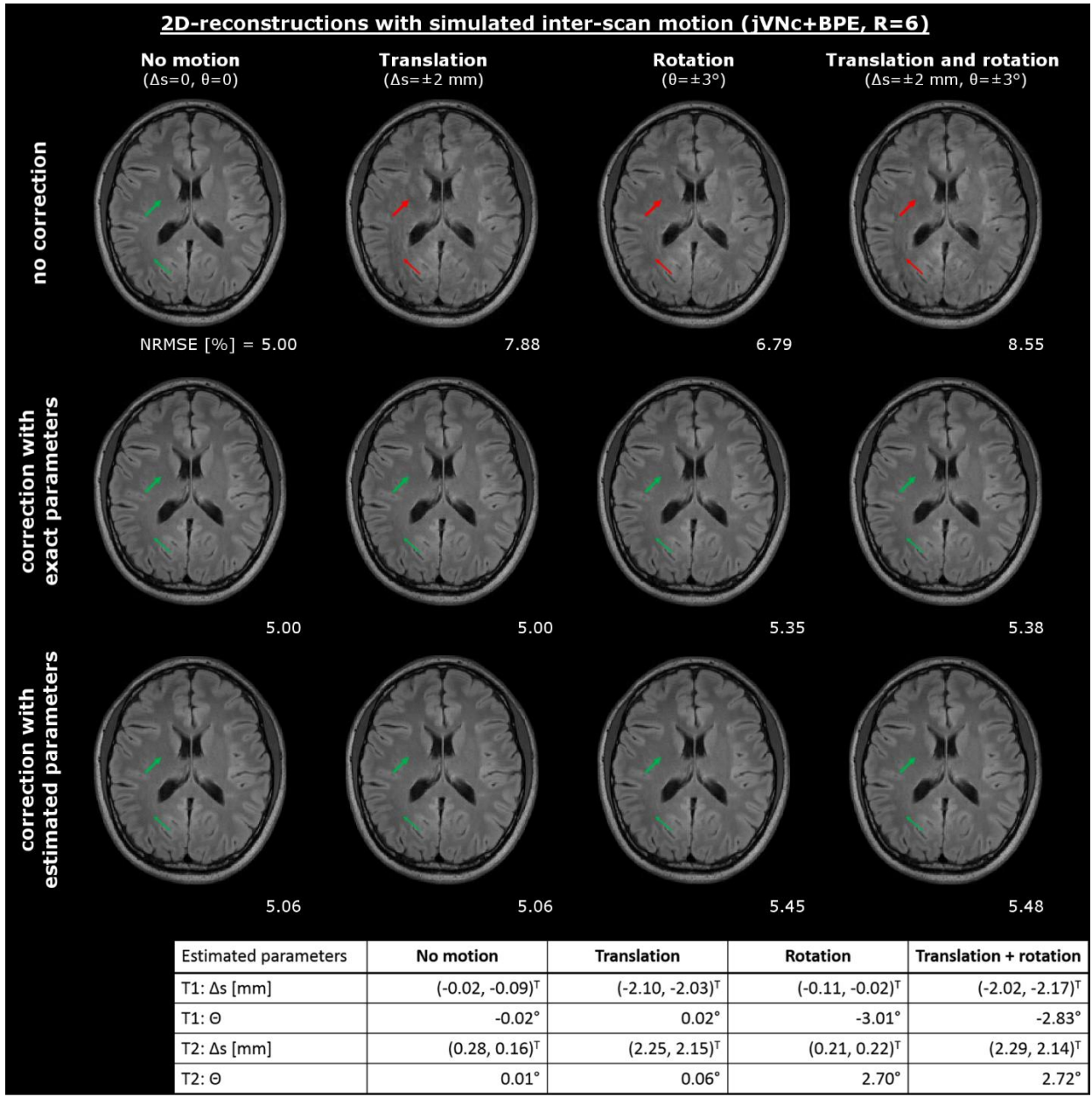

Figure 5: Inter-scan motion caused artifacts in our joint multi-contrast reconstruction (*jVNc+BPE*), which were significantly reduced using our motion mitigation technique. The performance was evaluated using both the exact and estimated motion parameters which were derived from the initial *SENSE* reconstructions and are reported in the bottom of the figure.

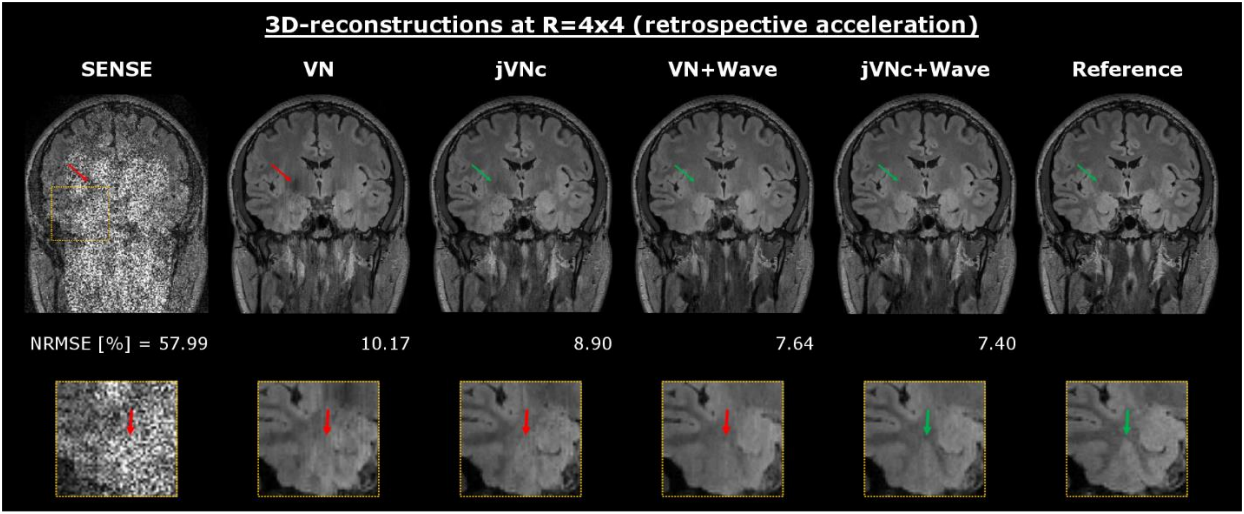

Figure 6: At *R*=4x4 acceleration, *VN* efficiently denoised the initial *SENSE* reconstruction but resulted in residual aliasing (thin arrow), striping artifacts and over-smoothing (fat arrow in zoom-in). This was improved in multi-contrast *jVNc*, however striping artifacts were only mitigated in the *Wave* reconstructions. The overall best performance was obtained by *jVNc+Wave*.

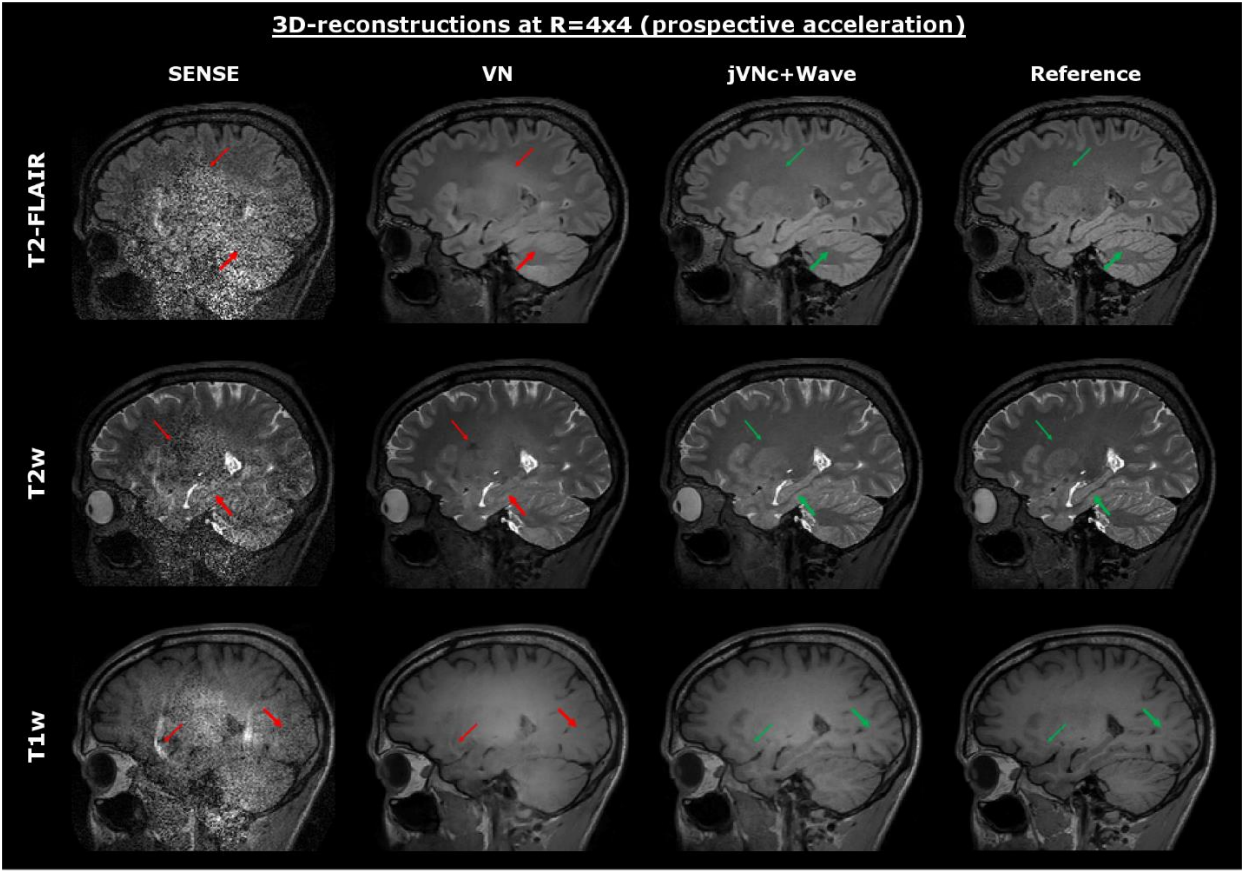

Figure 7: The variational networks were tested on prospectively accelerated data acquired at *R*=4x4 acceleration (combined TA=2:53 min). The sequence specific contrast was retained in all scans, but *jVNc+Wave* better preserved fine anatomical details (fat arrow) and exhibits fewer artifacts (thin arrow) than *VN*. Nevertheless, at such high acceleration (*R*=16) also the *jVNc+Wave* reconstructions resulted in small image blurring, for example in the cerebellum of T2-FLAIR.

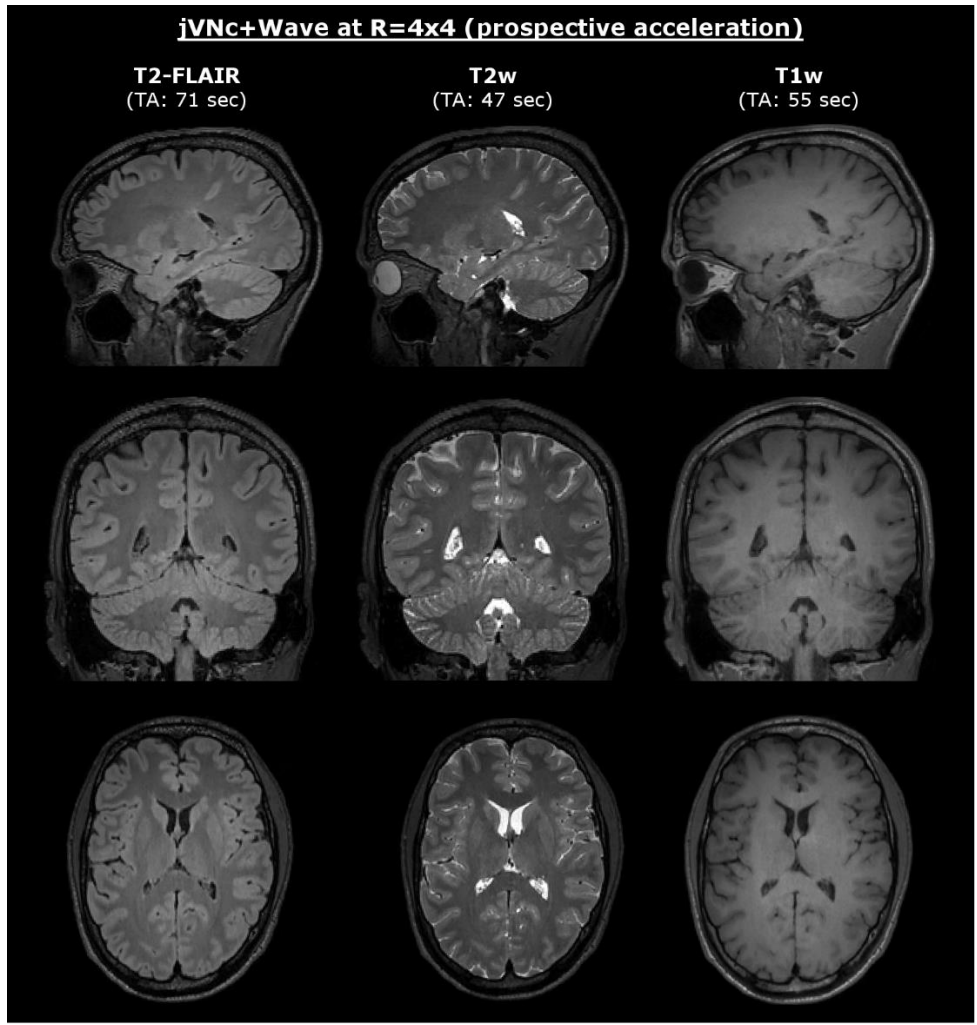

Figure 8: At 1 mm isotropic resolution T1w, T2w and T2-FLAIR were acquired at *R*=16-fold acceleration and reconstructed using *jVNc+Wave* (combined TA=2:53 min).